\documentclass[10pt,superscriptaddress,prl,twocolumn]{revtex4}
\usepackage{amsfonts}
\usepackage{amsmath}
\usepackage{amssymb}
\usepackage{graphicx}
\usepackage{graphics}
\usepackage[usenames]{color}

\begin{document}
\title{Rydberg Electrons in a Bose-Einstein Condensate}
\author{Jia Wang}
\affiliation{Department of Physics, University of Connecticut, Storrs, CT 06269, USA}
\author{Marko Gacesa}
\affiliation{Department of Physics, University of Connecticut, Storrs, CT 06269, USA}
\author{Robin C\^{o}t\'{e}}
\affiliation{Department of Physics, University of Connecticut, Storrs, CT 06269, USA}
\begin{abstract}
We investigate a hybrid system composed of ultracold Rydberg atoms immersed in an
atomic Bose-Einstein condensate (BEC). The coupling between the Rydberg electrons and
BEC atoms leads to the excitation of phonons, the exchange of which induces Yukawa interaction
between Rydberg atoms. Due to the small electron mass, the effective charge associated with this
quasi-particle-mediated interaction can be large, while its range is equal to the healing 
length of the BEC, which can be tuned by adjusting the scattering length of the BEC atoms. We find that for small healing lengths, the distortion of the BEC can ``image'' the wave function density of the Rydberg electron, while large healing lengths induce an attractive Yukawa potential between the two Rydberg atoms that can form a new type of ultra-long-range molecule. We discuss both cases for a realistic system.
 
\end{abstract}
\maketitle
Impurities in a Bose-Einstein condensate (BEC) have attracted much attention and 
motivated the investigation of a wide range of phenomena.
For example, the motion of a single impurity in a BEC can probe the superfluid 
dynamics \cite{Timmermans1998,KetterlePRL2000,PitaevskiiPRA2004}, while an ionic 
impurity in a BEC can form a mesoscopic molecular ion \cite{RobinPRL2002}. 
Due to the self-energy induced by phonons (excitations of the BEC), a neutral 
impurity can self-localize in both a homogeneous and a harmonically trapped 
BEC \cite{TimmermansPRA2006,BlumePRA2006,JakschEPL2008}, which sheds 
light on polaron physics \cite{TimmermansPRL2006,TimmermansPRA2013}. 
Exchanging phonons between multiple impurities induces an attractive Yukawa 
potential between each pair of impurities \cite{ViveritPRA2000,BijlsmaPRA2000}, 
which leads to the so called ``co-self-localization'' \cite{TimmermansJNP2011} and is 
related to forming bipolarons and multipolarons \cite{DevreesePRA2013}. 
Recent experiments, where an atom of a BEC is excited into a Rydberg 
state \cite{JonathanNature2013} to study phonon excitations and collective oscillations,
open the door to exploration of the electron-phonon coupling in ultracold degenerate gases, a
phenomenon responsible for the formation of Cooper pairs of two repelling electrons in BCS superconductivity \cite{BCStheory}. 

\begin{figure}[t]
\includegraphics[width=0.5 \textwidth]{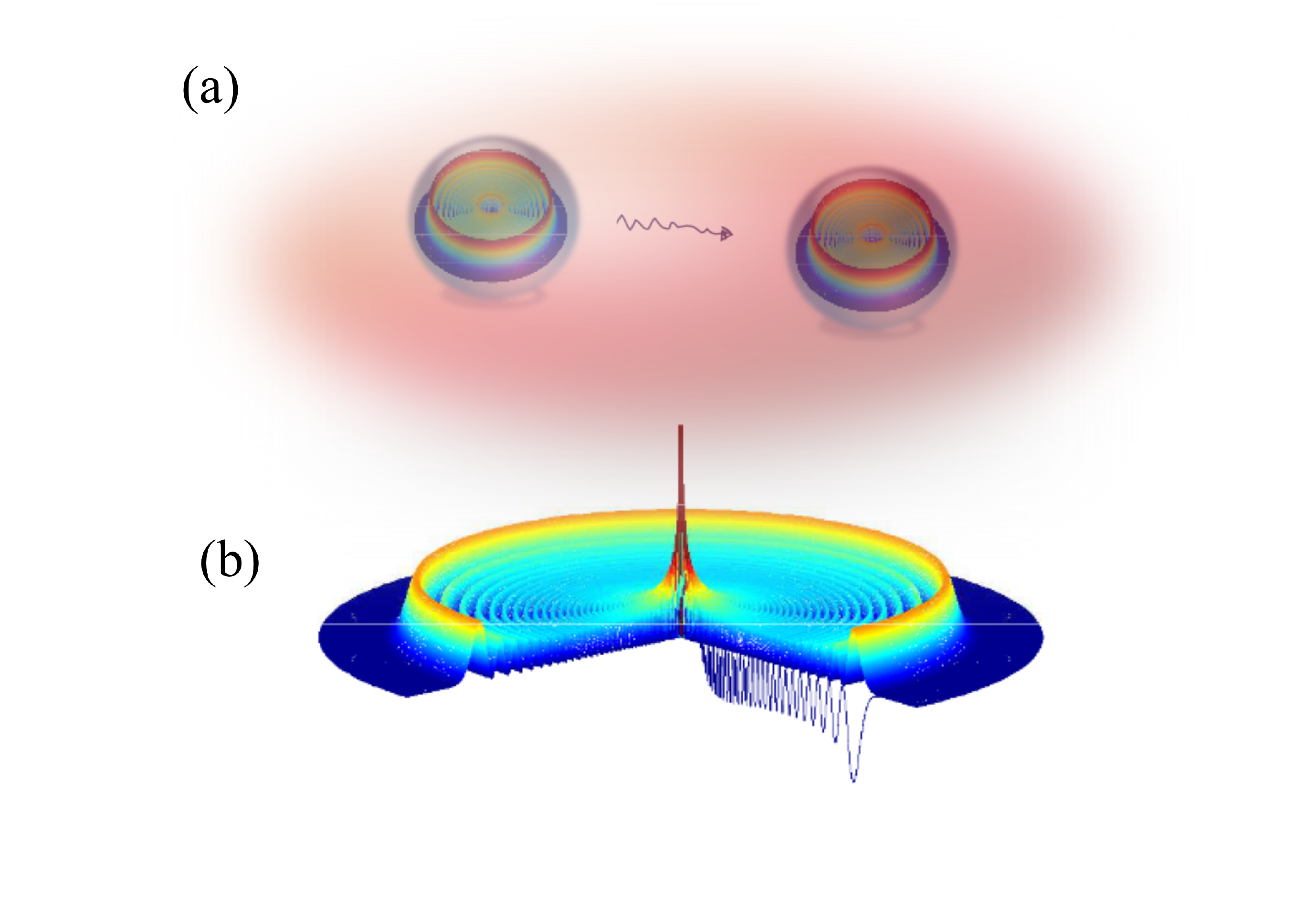}
\caption{(Color online) (a) Sketch: two Rydberg atoms immersed in an atomic BEC exchange phonons. The Rydberg electrons are represented by the surface plots inside the spheres, plotted in (b) along with the interaction potential curve within the $s$-wave approximation.}\label{RydBEC}
\end{figure}

In this Letter, we study Rydberg atoms immersed in a homogeneous BEC, as
sketched in Fig.~\ref{RydBEC}(a). Rydberg atoms consist of an ion core and a highly excited electron with its oscillatory wave function $\Psi_e$ extending to large distances of the order of
${\sim}n^2 a_0$ ($n$: principle quantum number, $a_0$: Bohr radius). 
As pointed out by Fermi \cite{Fermi1934}, the interaction between the quasi-free electron 
at $\mathbf{x}$ and a ground state atom at $\mathbf{r}$ can be approximated at 
low scattering energies by a contact interaction parametrized by an energy-dependent $s$-wave scattering length $A_s\left({k}\right)=-k^{-1}\tan \delta_s (k)$,
\begin{equation}
  V_s \left( {\mathbf{x},\mathbf{r}} \right) = \frac{2\pi \hbar^2}{m_e} A_s  \left[ {k\left( r \right)} \right] \delta ^{\left( 3 \right)} \left( {\mathbf{x} - \mathbf{r}} \right) \;.
  \label{eq:V_s}
\end{equation}
While the $s$-wave approximation is valid for qualitative analysis, we include higher-partial
wave contributions for quantitative results \cite{note-p-wave}.
$A_s \left( {k} \right)$ depends on the scattering energy via the local wave number 
$k ( r )$ given by
\begin{equation} 
   \frac{\hbar^2k (r )^2}{2m_e} = -\frac{R_y}{( {n - \delta _{\ell_e} })^2} 
   + \frac{e^2}{4\pi \epsilon _0 r} \; ,
\end{equation}
where $R_y$ is the Rydberg constant, $\epsilon _0$ the vacuum permittivity, $e$ and 
$m_e$ the charge and mass, respectively, of the electron with angular momentum 
$\ell_e$ and quantum defect $\delta_{\ell_e}$. For low-$\ell_e$ state, Eq.(\ref{eq:V_s}) 
gives an effective interaction between Rydberg and ground state atoms as
\begin{equation}
{V_R} ( {\bf r} ) \approx \frac{{2\pi {\hbar ^2}A_s\left[ {k\left( {{r}} \right)} \right]}}{{{m_e}}}{\left| {{\Psi _e}( {\bf r} )} \right|^2},
\end{equation}
which leads to an attraction and formation of ultra-long-range 
Rydberg molecules for $A_s<0$ \cite{GreenePRL2000}. The electron density and 
corresponding oscillatory potential 
are sketched in Fig.~\ref{RydBEC}(b) for a Rydberg $ns$ ($\ell_e=0$) state. 
High-$\ell_e$ states with negligible $\delta_{\ell_e}$ are nearly degenerate, and their
coupling gives electronic wave functions with strong quantum interference patterns.
For alkali metals ({\it e.g.}, Rb or Cs), these interactions are strong enough to support 
very extended bound states, usually referred to as ``trilobite states'', that possess 
a strong permanent dipole moment. The observation of ``trilobite-like states''
\cite{BendkowskyNature2009,LiWscience2011,BellosPRL2013,Andersonarxiv2014,Krupparxiv2014}, has motivated the studies of the $p$-wave electron (leading to ``butterfly states''
\cite{HamiltonJPB2002}), and Rydberg electrons scattering off a perturber with a permanent 
dipole moment \cite{RittenhousePRA2012}. 

In our system, Rydberg electrons interact with the coherent ground state of BEC, {\it i.e.} 
with many atoms, resulting in collective excitations described as phonons 
(scalar bosons). One of the most profound conceptual advances in physics is that the 
exchange of particles can produce a force ({\it e.g.}, the electromagnetic force is generated
by charges exchanging virtual photons). Exchanging phonons in a BEC will lead to a 
Yukawa potential. As we describe below, under appropriate conditions, we find two regimes.
For a BEC with a small healing length $\xi$, the Yukawa potential is short-ranged, 
and distorts the BEC locally, ``mapping" the electron density onto the BEC density. For a large $\xi$, the Yukawa potential is long-ranged and can bind Rydberg 
atoms and form a new type of ``ultra-long-range'' molecule.

We first consider a homogeneous BEC in the absence of impurities, described by the Hamiltonian
\begin{equation}
H_{\rm BEC}  = \sum\limits_\mathbf{k} {\frac{{\hbar ^2 k^2 }}{{2m_B}}c_\mathbf{k}^\dag  c_\mathbf{k}  + \frac{u_B}{{2\Omega_V}}\sum\limits_{\mathbf{kpq}} {c_\mathbf{k}^\dag  c_\mathbf{p}^\dag  c_\mathbf{q} c_{\mathbf{k} + \mathbf{p} - \mathbf{q}} } },
\label{eq:H_BEC}
\end{equation}
where $u_B=4\pi\hbar^2 a_{B}/m_B$ is the coupling constant between the atoms of mass
$m_B$ and scattering length $a_B$, $\Omega_V$ is the quantization volume, and 
$c_\mathbf{k}^\dag$ ($c_\mathbf{k}$) is the creation (annihilation) operator of bosonic 
atoms with momentum $\mathbf{k}$. If most atoms occupy the ground state ($\mathbf{k}$=0), one can replace $c_0^{\dag}$ and $c_0$ by the $c$-number $\sqrt{N_0}$ and expand Eq.(\ref{eq:H_BEC}) in the decreasing order of $N_0$. The number of
atoms is given by $N=N_0 + \sum_{{\bf k}\ne 0} c_{\bf k}^\dag  c_{\bf k} $. By keeping the terms 
of the order $\sqrt{N_0}$ or higher, $H_{\rm BEC}$ can be diagonalized via the Bogoliubov transformation 
$c_{\bf q}^\dag = u_q b_{\bf q}^\dag + v_q b_{-{\bf q}}$. The resulting effective Hamiltonian is 
${\cal H}_{\rm BEC} = \sum_{\bf q} \hbar \omega_q (b_{\bf q}^\dag  b_{\bf q} +1/2) $, 
where $\hbar \omega _q  = (\epsilon_q^2  + 2u_B \rho _B \epsilon_q)^{1/2}$, 
with $\epsilon_q = {\hbar ^2}{q^2}/ 2 m_B$ and the BEC number density 
$\rho_B=N/\Omega_V$. The Bogoliubov operator $b_{\bf q}^\dag$ ($b_{\bf q}$) creates
(annihilates) a quasi-particle (or phonon) of momentum $\bf q$ when applied to the
ground state $|0\rangle$: $b_{\bf q}^\dag | 0 \rangle  = | {\bf q} \rangle$. The local density operator $\hat \rho\left( \mathbf{r} \right) =\Omega_V^{-1}
\sum\limits_{\mathbf{p},\mathbf{q}} {e^{i\mathbf{q}\cdot\mathbf{r}} 
c_{\mathbf{p} + \mathbf{q}}^\dag  c_\mathbf{p} } $ can be written as,
\begin{equation}
\hat \rho\left(\mathbf{r} \right) \approx \frac{{{N_0}}}{\Omega_V} + \frac{{\sqrt {{N_0}} }}{\Omega_V}\sum\limits_{\mathbf{q} \ne 0} {{e^{i\mathbf{q} \cdot \mathbf{r}}}\left( {{u_\mathbf{q}} + {v_\mathbf{q}}} \right)} \left( {b_\mathbf{q}^\dag  + {b_{ - \mathbf{q}}}} \right) ,
\end{equation}
and the interaction between a Rydberg electron and BEC atoms 
$H_{\rm INT} = \int d^3 r \rho ({\bf r})V_R( {\bf r} )$ as
\begin{equation}
{H_{\rm INT}} \approx \frac{{{N_0}}}{\Omega_V}{V_0}+\frac{{\sqrt {{N_0}} }}{\Omega_V}\sum\limits_{\mathbf{q} \ne 0} {\left( {{u_\mathbf{q}} + {v_\mathbf{q}}} \right)\left( {b_\mathbf{q}^\dag  
+ {b_{ - \mathbf{q}}}} \right)V_{\mathbf q}},
\end{equation}
where $V_{\bf q} = \int d^3 r V_R({\bf r})e^{i{\bf q} \cdot {\bf r}}$ is the Fourier transform 
of the potential. 
Applying the perturbation theory gives the first and second order corrections to the 
ground state energy: $E^{\left(1\right)} = \int d^3 r \rho _B V_R( {\bf r})$ and
\begin{equation}\label{perturbedenergy2}
E^{\left(2\right)} = -\frac{\rho _B m_B}{2\pi \hbar ^2} \int d^3 r d^3 r' V_R ({\bf r} )
\frac{e^{ - | {\bf r} - {\bf r}' |/\xi}}{| {\bf r} - {\bf r}' |}V_R ( {\bf r} ) ,
\end{equation}
by taking the thermal limit of $V^{- 1} \sum_{\bf q}  \to  ( 2\pi )^{- 3} \int d^3 q$ and integrating over $\bf q$, where $\rho_B$ is assumed to be a constant. Note that $N_0$ can be replaced by the total atom number $N$ at this level of approximation.

Under approximation in Eq.~(\ref{eq:V_s}), $E^{(1)}=2\pi \rho _B \hbar^2 {\bar a}_e/m_e$
is the mean-field energy shift given in terms of the average scattering length
${\bar a}_e = \int d^3 r A_s [ k( r)]|\Psi _e( {\bf r} )|^2$, 
while $E^{(2)} \approx   \int d^3 r d^3 r' |\Psi _e ( {\bf r} ) |^2
V_Y ({\bf r}-{\bf r}' )|\Psi _e ( {\bf r}' ) |^2/2$ involves the Yukawa potential
\begin{equation}\label{SingleYukawa}
V_Y\left({\mathbf{r}-\mathbf{r}'}\right) = -{{\tilde Q}^2} \frac{e^{ - | {\bf r} - {\bf r}' |/\xi }}
{| {\bf r} - {\bf r}' |},
\end{equation}
where its range $\xi=1/\sqrt{16\pi \rho_B a_B}$ is exactly equal to the BEC 
healing length, and $\tilde Q^2 \approx 4\pi \hbar ^2 \bar a_e^2 \rho _B m_B / m_e^2$ 
characterizes its strength; the ``effective charge'' $\tilde Q$ emphasizes the analogy 
with Coulomb interactions. The term ${E^{\left( 2 \right)}}$ can be understood as the self-interaction 
of electrons by a Yukawa potential induced via phonon exchange at two different positions. 
This term is crucial in studies of self-localization of impurities in a BEC. However, in our system, 
the Rydberg electrons are already localized by strong Coulomb forces with ion cores. Therefore, 
the distorted BEC density, under appropriate conditions, can reflect the oscillatory 
nature of $\Psi _e$ and ``image'' the Rydberg electron. 

\begin{figure}[t]
\includegraphics[width=0.45 \textwidth]{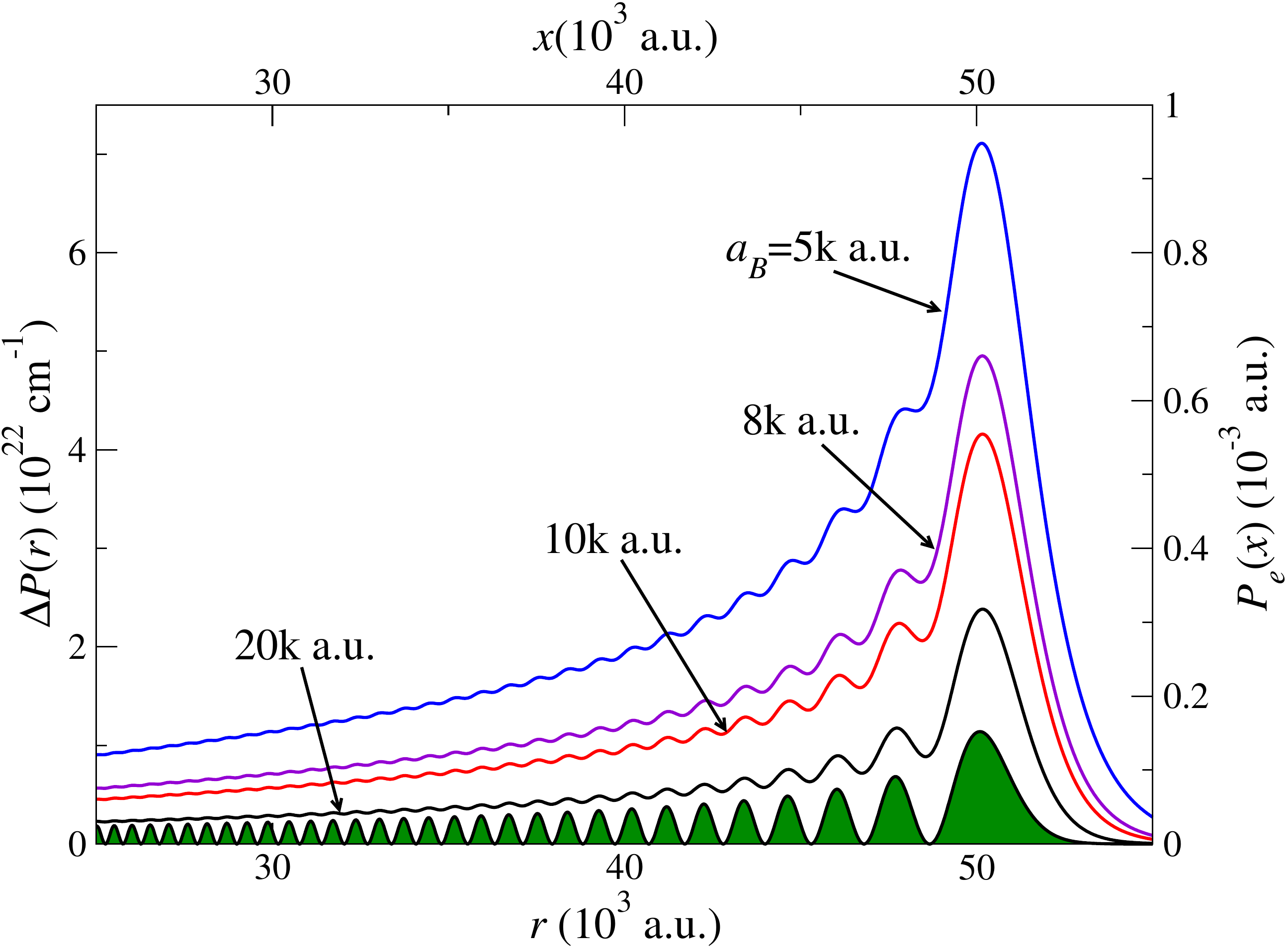}
\caption{(Color online)  Comparison of radial electron probability density $P_e\left({x}\right)=4\pi x^2|\Psi_e\left({x}\right)|^2$ (green filled curve)and BEC local density 
distortion $\Delta P\left({r}\right)=4\pi r^2\delta\rho (r)$ (solid curves) for a $^{87}$Rb BEC with density 
$\rho_B=2\times10^{13}$ cm$^{-3}$ and various scattering length $a_B$ (in units of k=$10^3$ $a_0$) as indicated.} \label{density3D}
\end{figure}

\begin{figure}[t]
\includegraphics[width=0.45 \textwidth]{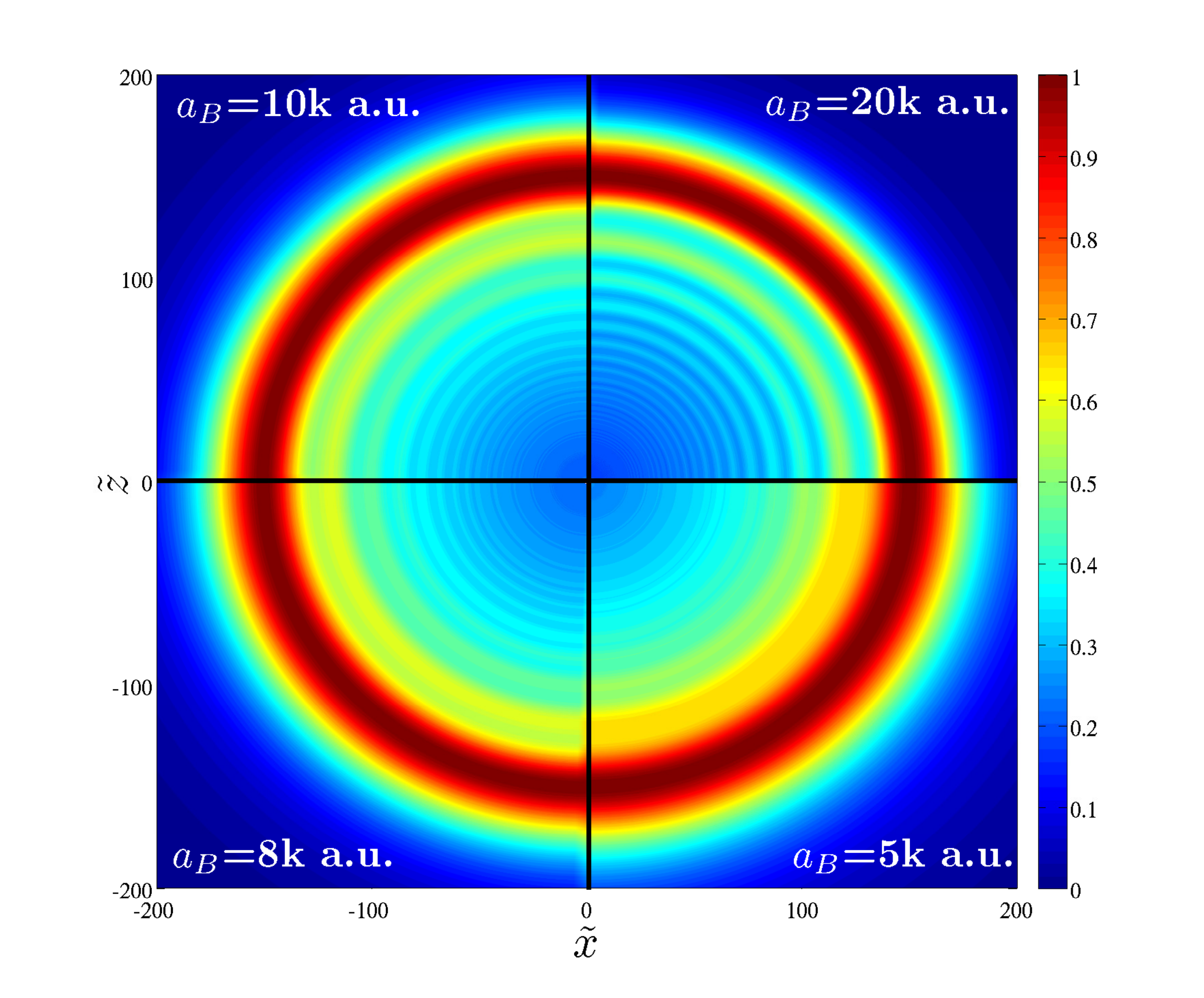}
\caption{(Color online)  The BEC density distortion $\Delta P_{2D}/\max\left({\Delta P_{2D}}\right)$ in the x-z plane, where $\Delta P_{2D}=2\pi r \delta\rho$. To better illustrate the oscillatory behavior at large distances, we use an exponential scale: $\tilde x=\exp(x/l_0)$ and $\tilde z=\exp(z/l_0)$, where $l_0=1$k a.u.}\label{density2D}
\end{figure}

To the first order, the perturbed ground state given by 
$\left| {\tilde 0} \right\rangle  = \left| 0 \right\rangle  - \left({{\sqrt {{N_0}} }/\Omega_V}\right)\sum_{\mathbf{q} \ne 0} {\left( {{u_\mathbf{q}} + {v_\mathbf{q}}} \right){V_\mathbf{q}}/\left({\hbar\omega_q}\right)\left| \mathbf{q} \right\rangle }$
leads to the BEC density distortion $\delta \rho \left( \mathbf r \right) \equiv \left\langle {\hat \rho \left( \mathbf r \right)} \right\rangle  - {\rho _B}$
\begin{equation}
\frac{\delta \rho ( {\bf r} )}{\rho _B} 
=  - \frac{m_B}{\hbar ^2 \pi}\int d^3 r' V_R ( {\bf r}' )
\frac{e^{ -| {\bf r} - {\bf r}' |/\xi }}{| {\bf r} - {\bf r}' |}.
\label{eq9}
\end{equation}
Eq. (\ref{eq9}) shows that $\delta \rho ( {\bf r})$ is affected by ``averaging'' the effective interaction $V_R$ within the range $\xi$. The oscillatory nature of $\Psi_e$ can be imaged onto $\delta \rho ( {\bf r})$ \cite{PfauArxiv2014}. However, if $\xi$ is larger than the local wavelength of the Rydberg electron, the averaging will erase this signature. This effect is illustrated in Fig.~\ref{density3D} for a $^{87}$Rb$(160s)$ Rydberg atom in a $^{87}$Rb BEC with $\rho_B=2\times10^{13}$ 
cm$^{-3}$, by comparing the radial probability density $P_e\left({x}\right)=4\pi x^2|\Psi_e\left({x}\right)|^2$ with $\Delta P\left({r}\right)=4\pi r^2 \delta \rho \left({r}\right)$ for different scattering lengths $a_B$. Larger values of $a_B$ produce a sharper signature of the
oscillation, albeit an overall smaller distortion amplitude. This effect is better illustrated in a 2D density plot (Fig.~\ref{density2D}), where different quadrants represent the ``normalized'' 2D distortion densities $\Delta P_{2D}/\max\left({\Delta P_{2D}}\right)$ in the $x$-$z$ plane for four different $a_B$. Here, $\Delta P_{2D}=2\pi r \delta\rho$, and $x$ and $z$ are rescaled by $\tilde x=\exp(x/l_0)$ and $\tilde z=\exp(z/l_0)$, where $l_0=1$k a.u., so that the effects at large distances are emphasized. It is evident that the oscillations for $a_B=5$k a.u. (fourth quadrant) are much blurrier than for $a_B=20$k a.u. (first quadrant).

For a large healing length $\xi$, the averaging of $V_R$ masks the effect of the electron
self-interaction due to the phonon exchange. However, the phonon exchange 
still mediates non-trivial interactions between the Rydberg atoms. Without a BEC, two 
Rydberg atoms experience strong long-range interactions, leading to formation of macrodimers \cite{macrodimer-1} and the interaction
blockade \cite{LukinPRL2001,TongPRL2004,SingerPRL2004,VogtPRL2006,LiebischPRL2005}. For two $ns$ Rydberg atoms separated by R, this interaction
is repulsive with its leading term being the van-der-Waals (vdW)
$+C_6/R^6$ term, where $C_6 \propto n^{11}$ \cite{macrodimer-2}. 
Immersed in a BEC, however, the exchange of phonons between two Rydberg atoms 
gives rise to a Yukawa potential. We derive this potential within the Born-Oppenheimer 
(BO) approximation, starting from the interaction of two Rydberg atoms, located at 
$\mathbf{R}_1$ and $\mathbf{R}_2$, and BEC atoms (after applying the Bogoliubov transformation)
\begin{eqnarray}
 H_{\rm INT} \approx \frac{N_0}{\Omega_V}  {^1\cal V}_{0}+\frac{N_0}{\Omega_V}
 {^2\cal V}_{0}+\frac{\sqrt{N_0}}{\Omega_V}\sum\limits_{\mathbf{q} \ne 0} (u_q+v_q ) \nonumber \\
 \times ( b_\mathbf{q}^\dag   + b_{-\mathbf{q}} ) ({^1\cal V}_{\mathbf{q}} 
 e^{i\mathbf{q}\cdot\mathbf{R}_1} + {^2\cal V}_{\mathbf{q}}e^{i\mathbf{q}\cdot\mathbf{R}_2} ).
\end{eqnarray}
Here, ${^i\cal V}_\mathbf{q}  \equiv \int d^3 r V_i ( \mathbf{r} )e^{i\mathbf{q}\cdot\mathbf{r}}$, 
where $V_i ( \mathbf{r} )$ describes the interaction of ``impurity" $i$ and the BEC atoms 
in coordinate space. Within perturbation theory, the first order correction $E^{(1)}$ gives 
a mean-field energy shift similar to the single Rydberg atom case in the thermal limit. 
For spherically symmetric interactions (where ${^i\cal V}_\mathbf{q}={^i\cal V}_\mathbf{-q}
={^i\cal V}_q$ is real), the second order correction is 
\begin{equation}
   E^{( 2 )} =  - \frac{\rho _B}{( 2\pi )^3} \int d^3 q \frac{A_q + 2 B_q  e^{i q R \cos \theta_q}}
    {\epsilon_q + 2 u_B \rho _B},
\end{equation}
where $A_q= ({^1\cal V}_q)^2+ ({^2\cal V}_q)^2$, $B_q={^1\cal V}_q \cdot {^2\cal V}_q$, $R=|\mathbf{R}_1-\mathbf{R}_2|$ is the Rydberg atoms separation, and $\theta_q$ is the angle between the vector $\mathbf{R}$ and $\mathbf{q}$. The term 
containing $A_q$ can be understood as the self-localizing energy for both Rydberg atoms 
calculated previously, and will be neglected together with the mean-field energy shift $E^{(1)}$
for the study of relative dynamics of Rydberg atoms, since they simply contribute a constant energy shift. The term containing $B_q$ leads to the BO potential
\begin{equation}\label{IBinteraction}
   U ( R ) = - \frac{\rho _B}{( 2\pi )^3} \int d^3 q \frac{2 B_q e^{i q R \cos \theta_q}}
    {\epsilon_q + 2 u_B \rho _B},
\end{equation} which can be easily generalized to interactions between any two impurities immersed in a BEC \cite{BijlsmaPRA2000}.

The BO approach allows the study of the adiabatic corrections induced by the motion of 
Rydberg atoms. The diagonal adiabatic correction 
$\Delta E^{( 2 )}  =   \hbar ^2  \langle {\tilde 0} |\mathord{\buildrel{\lower3pt\hbox{$\scriptscriptstyle\leftarrow$}}
\over \partial } _R \vec \partial _R | {\tilde 0} \rangle /m_I$ depends on the impurity 
mass $m_I$, where $\left| {\tilde 0} \right\rangle  = \left| 0 \right\rangle  - \sum_{\mathbf{q} \ne 0} {\frac{{\left\langle \mathbf{q} \right| H_{\mathrm{INT}} \left| 0 \right\rangle }}{{\hbar \omega _q }}} \left| \mathbf{q} \right\rangle$ is the perturbed ground state. 
Therefore, $\Delta E^{\left( 2 \right)}  = \frac{{\hbar ^2 }}{{m_I }}\sum_{\mathbf{q} \ne 0} {\frac{{\left\langle 0 \right|\partial _R H_{\mathrm{INT}} \left| \mathbf{q} \right\rangle \left\langle \mathbf{q} \right|\partial _R H_{\mathrm{INT}} \left| 0 \right\rangle }}{\hbar ^2 \omega _q^2 }}$, so that, in the thermal limit and neglecting the constant energy shift terms, the diabatic correction to 
$U\left( R \right)$ is 
\begin{equation}\label{IBcorrection}
\Delta U( {R} ) =  - \frac{\hbar ^2 }{2m_I }\frac{\rho _B}{(2\pi )^3} \int d^3 q 
\frac{B_q q^2 \cos ^2 \theta _q e^{i q R \cos \theta_q}}
       {\sqrt {\epsilon_q ( \epsilon_q + 2u\rho _B )^3 } }.
\end{equation}
The $e^{i q R \cos \theta_q}$ term in Eqs. (\ref{IBinteraction}) and (\ref{IBcorrection}) implies that, for a large $R$, only the terms where $q$ is small are important, leading to the asymptotic behavior of the potential
$U\left( {R} \right) \rightarrow  - \tilde Q^2 e^{ - R/\xi}/R$, 
where $\tilde{Q}^2 \approx \rho_B m_B B_{q=0}/\pi$. For the $s$-wave, 
the effective charge is $\tilde Q^2 \approx 4\pi \hbar ^2 \bar a_e^2 \rho _B m_B / m_e^2$.
Not surprisingly, we obtain the same Yukawa potential as in Eq. (\ref{SingleYukawa}), 
since phonon-exchange mediates the interaction. Note that $\tilde Q$ is inversely 
proportional to $m_e$ for Rydberg atoms, since the electrons are really the perturbers, as 
opposed to more massive neutral impurities for which $\tilde Q$ is inversely proportional to 
$m_I$. Hence, the induced interaction is much stronger for Rydberg atoms.
Under the same approximation, the adiabatic correction is given by $\Delta U\left( {R} \right) \rightarrow \left( {m_B /m_I} \right) (\tilde{Q}^2 /2 \xi )F\left( {R/\xi} \right)$,
with $F\left(x\right)$ defined as
$F ( x ) =   \frac{2}{\pi } - \frac{4}{\pi x^2 } - \frac{2 f( - 1,x )}{x^2} + \frac{f(0,x)}{x} - f(1,x)$,
where $f(n,x)=I_n (x)- L_n (x)$ is given in terms of the modified Bessel function of the first kind 
$I_n (x)$ and the modified Struve function $L_n (x)$. The asymptotic behaviors are 
$F(R/\xi )\rightarrow 4/(3\pi)$ for $R \ll \xi$, and $F(R/\xi)\rightarrow 12\xi^4/(\pi R^4)$ 
for $R \gg \xi$. We note that for $R\gg \xi$, these imply a vanishing adiabatic potential 
$U\left({R}\right)$ so that the adiabatic correction becomes dominated by a repulsive $1/R^4$ term. 
As expected, the diabatic correction can be neglected if $m_B\ll m_I$. 
A surprising limit is reached for a very large healing length $\xi$, 
which can be achieved by using a Feshbach resonance to tune $a_B \rightarrow 0$.
Then, the diabatic correction also vanishes:
$\lim_{\xi  \to \infty } \Delta U\left( R \right) = 0$, even when $m_B$ is larger than $m_I$. Notice that, in this limit, $U\left({R}\right)$ reduces to the Coulomb potential.

\begin{figure}[t]
\includegraphics[width=0.45 \textwidth]{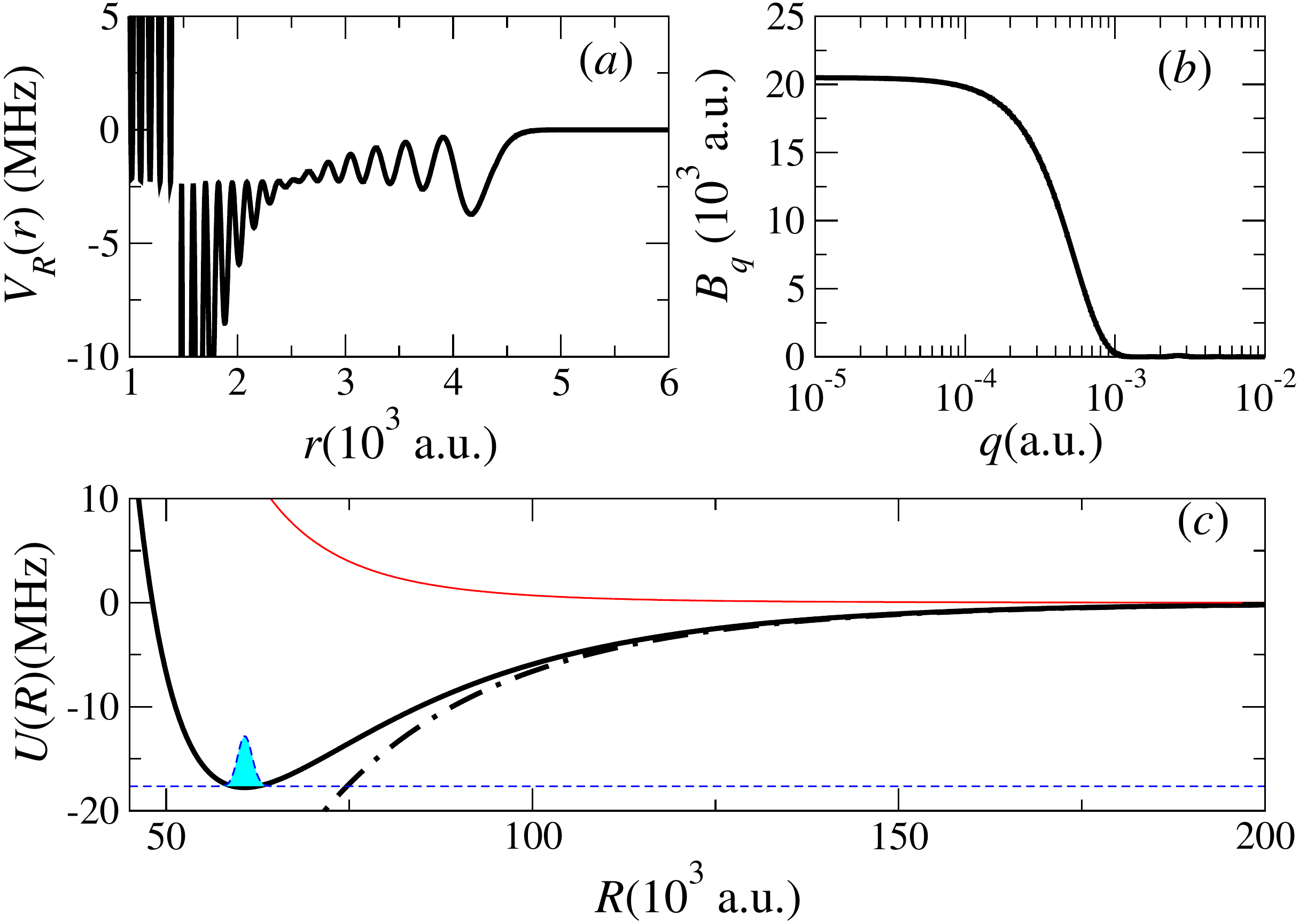}
\caption{(Color online) (a) Interaction potential between a Rydberg $^{87}$Rb$(50s)$ atom and a ground state $^{87}$Rb atom. (b) $B_q$ as a function of $q$. (c) Effective potential between two Rydberg $^{87}$Rb$(50s)$ atoms in a BEC. The tail is dominated by the Yukawa potential (dash-dotted curve) while the blue filled curve represents the lowest bound state.} \label{RydInt}
\end{figure}

To illustrate these predictions, we consider a realistic system of two $^{87}$Rb$(50s)$ Rydberg
atoms immersed in a BEC of $^{87}$Rb atoms of density $\rho_B=10^{13}$ cm$^{-3}$. 
To ensure a healing length $\xi$ much larger than the Rydberg atoms, the scattering length 
between the BEC atoms is tuned to $a_B=10$ a.u. ({\it e.g.}, via a Feshbach resonance), so that
$\xi = 3.66 \times 10^4$ a.u. The numerical ``trilobite-like'' interaction shown in 
Fig.~\ref{RydInt}($a$) is constructed using the first-order perturbative model 
\cite{Fermi1934,Omont1977} including $s$- and $p$- contributions \cite{note-p-wave},
with zero-energy scattering lengths, $A_{s}=-16.05$ $a_0$ and $A_{p}=-21.15$ $a_0$,
respectively \cite{BendkowskyPRL2010}. The states in the range $n=47-51$ were included and the resulting Hamiltonian diagonalized to obtain the $50s$ eigenstate \cite{note-ns}. 
Fig.~\ref{RydInt}(b) shows that $B_q$ computed using this potential converges 
to a constant $B_{q=0} \approx 2\times10^4$ a.u. for a small $q$, yielding 
an effective charge $\tilde Q^2 \approx 1.537\times10^{-3}$ a.u. 
Fig.~\ref{RydInt}(c) depicts the effect of immersing Rydberg atoms in a BEC: without the BEC,
two Rydberg atoms interact via the vdW potential $-C_6/R^6-C_8/R^8-C_{10}/R^{10}$
(repulsive solid-red curve with $C_6=-1.074\times10^{20}$, $C_8=7.189\times10^{26}$, 
and $C_{10}=-7.162\times10^{33}$, in a.u. for $^{87}$Rb($50s$) \cite{SingerJPB2005}.) 
In a BEC, the BO potential (solid black curve) is attractive at large separations, in agreement
with the Yukawa potential $-\tilde Q^2 e^{-R/\xi}/R$ (dash-dotted curve) at large distances,
before becoming repulsive at shorter range where the ``bare'' repulsive vdW 
interaction dominates the phonon-exchange contribution. The well produced by
the phonon-exchange can support bound levels; in the example above, its depth is
about -17.77 MHz, while the equilibrium separation is about 60$k$ a.u. (much larger
than the 5$k$ a.u. extension of the ``trilobite-like'' potentials). The large mass of Rb 
atoms leads to many bound levels; the three lowest are at about -17.64 MHz, -17.56 MHz 
and -17.47 MHz. The ground state wave function with a spatial width about 2$k$ a.u. is also shown in Fig.~\ref{RydInt}($c$),

\begin{figure}[t]
\includegraphics[width=0.45 \textwidth]{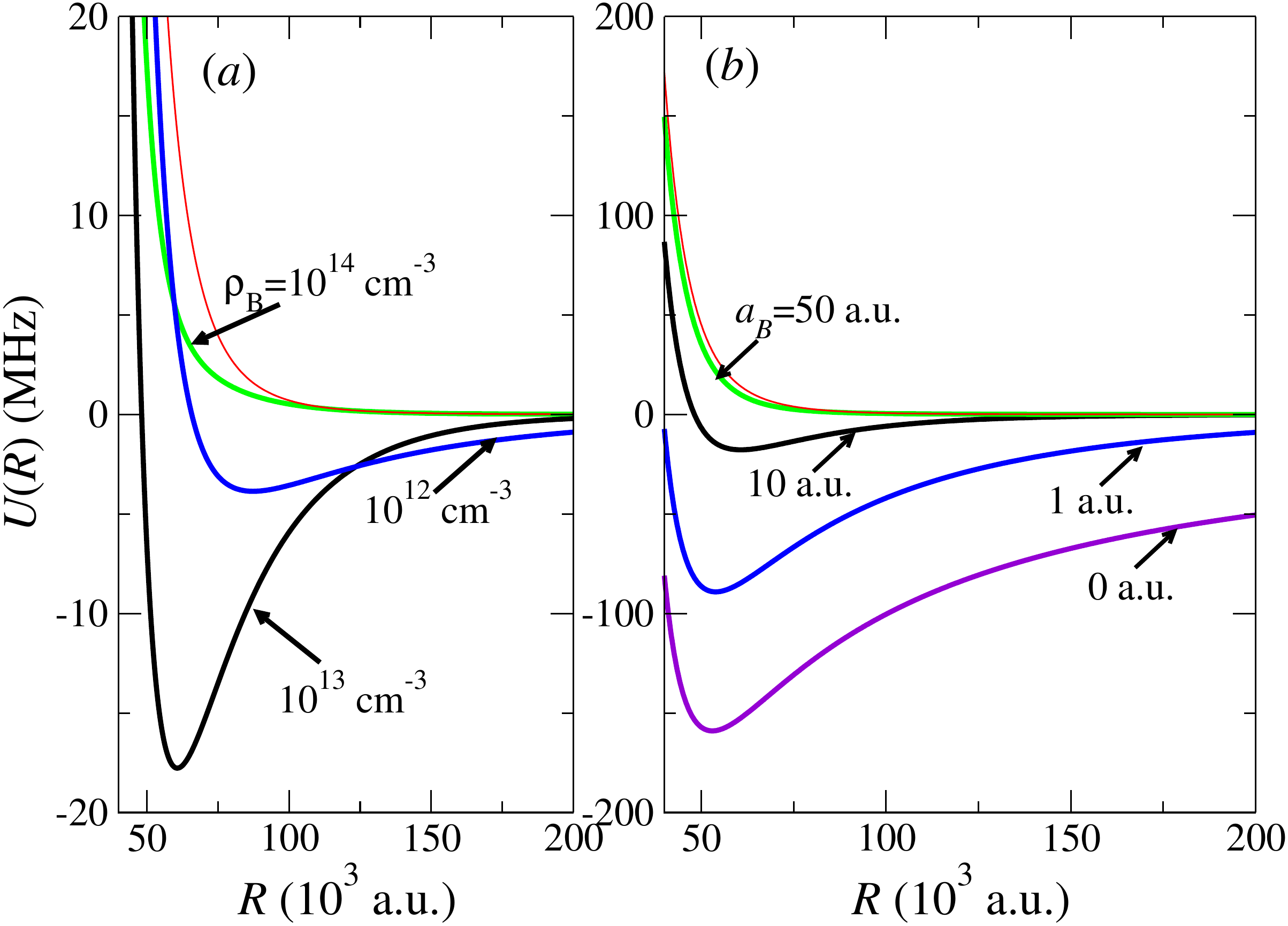}
\caption{(Color online) The thin solid red curves show the dispersion interactions between two $^{87}$Rb$(50s)$ atoms in a vacuum, while the thick curves show the interaction between these two Rydberg atoms immersed in a BEC with (a) fixed $a_B=10$ a.u. but different density $\rho_B$ and (b) fixed $\rho_B=10^{13}$ cm$^{-3}$ but different scattering length $a_B$.}\label{Upara}
\end{figure}

These results show how phonon-exchange modifies an otherwise repulsive interaction 
into a potential well capable of binding two Rydberg atoms. Fig.~\ref{Upara} explores the sensitivity
of the BO potentials to variations in density $\rho_B$ and scattering length $a_B$, and compares
them to the ``bare'' case (without BEC). The behavior of the BO curves can be 
understood qualitatively from the $s$-wave approximation with average scattering 
length $\bar a_e$: $\tilde Q^2$ is proportional to $\rho_B$ and $\xi$ to $\rho_B^{-1/2}$. 
The competition between the two effects as $\rho_B$ varies leads to a deeper BO curve 
for a moderate density (see Fig.~\ref{Upara}(a)). However, $\tilde Q^2$ is independent 
of $a_B$ while $\xi$ is proportional to $a_B^{-1/2}$, giving deeper BO curves as $a_B$ 
gets smaller (see Fig.~\ref{Upara}(a)). Hence, the BEC-induced interaction can be 
conveniently controlled by tuning $a_B$ via a Feshbach resonance;
in the limit $a_B=0$, the long-range Yukawa potential becomes an attractive Coulomb potential.

In summary, we studied BEC-induced interactions involving Rydberg impurities due to 
phonon-exchange, and found two limiting cases depending on the healing length $\xi$ of 
the BEC. For a small $\xi$, the BEC modulation can be used to ``image'' the wave function 
of the Rydberg electron, while large $\xi$ leads to the formation of ultra-long-range diatomic molecules. By tuning $a_B$, ``synthetic'' Coulomb potentials 
can be generated between neutral particles and their sign can be modified by using different Rydberg states for the two impurity atoms.
This long range interaction is well-behaved and easily controlled and, hence, 
opens promising avenues of research.  For example, for systems containing many 
Rydberg atoms impurities, this interaction might lead to 
crystallization \cite{RobertsPRL2009}, and be used to study the phase 
diagram of Yukawa bosons \cite{OsychenkoPRA2012}.

This work was partially supported by the U.S. Department of Energy, Office of Basic Energy Sciences (J.W.), the Army Research Office Grant No. W911NF-13-1-0213 (M.G.), and the National Science Foundation Grant No. PHY 1101254 (R.C.)

\bibliographystyle{apsrev}
\bibliography{refs}

\begin{thebibliography}{39}
\expandafter\ifx\csname natexlab\endcsname\relax\def\natexlab#1{#1}\fi
\expandafter\ifx\csname bibnamefont\endcsname\relax
  \def\bibnamefont#1{#1}\fi
\expandafter\ifx\csname bibfnamefont\endcsname\relax
  \def\bibfnamefont#1{#1}\fi
\expandafter\ifx\csname citenamefont\endcsname\relax
  \def\citenamefont#1{#1}\fi
\expandafter\ifx\csname url\endcsname\relax
  \def\url#1{\texttt{#1}}\fi
\expandafter\ifx\csname urlprefix\endcsname\relax\def\urlprefix{URL }\fi
\providecommand{\bibinfo}[2]{#2}
\providecommand{\eprint}[2][]{\url{#2}}

\bibitem[{\citenamefont{Timmermans and C\^ot\'e}(1998)}]{Timmermans1998}
\bibinfo{author}{\bibfnamefont{E.}~\bibnamefont{Timmermans}} \bibnamefont{and}
  \bibinfo{author}{\bibfnamefont{R.}~\bibnamefont{C\^ot\'e}},
  \bibinfo{journal}{Phys. Rev. Lett.} \textbf{\bibinfo{volume}{80}},
  \bibinfo{pages}{3419} (\bibinfo{year}{1998}).

\bibitem[{\citenamefont{Chikkatur et~al.}(2000)\citenamefont{Chikkatur,
  G\"orlitz, Stamper-Kurn, Inouye, Gupta, and Ketterle}}]{KetterlePRL2000}
\bibinfo{author}{\bibfnamefont{A.~P.} \bibnamefont{Chikkatur}},
  \bibinfo{author}{\bibfnamefont{A.}~\bibnamefont{G\"orlitz}},
  \bibinfo{author}{\bibfnamefont{D.~M.} \bibnamefont{Stamper-Kurn}},
  \bibinfo{author}{\bibfnamefont{S.}~\bibnamefont{Inouye}},
  \bibinfo{author}{\bibfnamefont{S.}~\bibnamefont{Gupta}}, \bibnamefont{and}
  \bibinfo{author}{\bibfnamefont{W.}~\bibnamefont{Ketterle}},
  \bibinfo{journal}{Phys. Rev. Lett.} \textbf{\bibinfo{volume}{85}},
  \bibinfo{pages}{483} (\bibinfo{year}{2000}).

\bibitem[{\citenamefont{Astrakharchik and
  Pitaevskii}(2004)}]{PitaevskiiPRA2004}
\bibinfo{author}{\bibfnamefont{G.~E.} \bibnamefont{Astrakharchik}}
  \bibnamefont{and} \bibinfo{author}{\bibfnamefont{L.~P.}
  \bibnamefont{Pitaevskii}}, \bibinfo{journal}{Phys. Rev. A}
  \textbf{\bibinfo{volume}{70}}, \bibinfo{pages}{013608}
  (\bibinfo{year}{2004}).

\bibitem[{\citenamefont{C\^ot\'e et~al.}(2002)\citenamefont{C\^ot\'e,
  Kharchenko, and Lukin}}]{RobinPRL2002}
\bibinfo{author}{\bibfnamefont{R.}~\bibnamefont{C\^ot\'e}},
  \bibinfo{author}{\bibfnamefont{V.}~\bibnamefont{Kharchenko}},
  \bibnamefont{and} \bibinfo{author}{\bibfnamefont{M.~D.} \bibnamefont{Lukin}},
  \bibinfo{journal}{Phys. Rev. Lett.} \textbf{\bibinfo{volume}{89}},
  \bibinfo{pages}{093001} (\bibinfo{year}{2002}).

\bibitem[{\citenamefont{Sacha and Timmermans}(2006)}]{TimmermansPRA2006}
\bibinfo{author}{\bibfnamefont{K.}~\bibnamefont{Sacha}} \bibnamefont{and}
  \bibinfo{author}{\bibfnamefont{E.}~\bibnamefont{Timmermans}},
  \bibinfo{journal}{Phys. Rev. A} \textbf{\bibinfo{volume}{73}},
  \bibinfo{pages}{063604} (\bibinfo{year}{2006}).

\bibitem[{\citenamefont{Kalas and Blume}(2006)}]{BlumePRA2006}
\bibinfo{author}{\bibfnamefont{R.~M.} \bibnamefont{Kalas}} \bibnamefont{and}
  \bibinfo{author}{\bibfnamefont{D.}~\bibnamefont{Blume}},
  \bibinfo{journal}{Phys. Rev. A} \textbf{\bibinfo{volume}{73}},
  \bibinfo{pages}{043608} (\bibinfo{year}{2006}).

\bibitem[{\citenamefont{Bruderer et~al.}(2008)\citenamefont{Bruderer, Bao, and
  Jaksch}}]{JakschEPL2008}
\bibinfo{author}{\bibfnamefont{M.}~\bibnamefont{Bruderer}},
  \bibinfo{author}{\bibfnamefont{W.}~\bibnamefont{Bao}}, \bibnamefont{and}
  \bibinfo{author}{\bibfnamefont{D.}~\bibnamefont{Jaksch}},
  \bibinfo{journal}{Europhys. Lett.} \textbf{\bibinfo{volume}{82}}
  (\bibinfo{year}{2008}).

\bibitem[{\citenamefont{Cucchietti and Timmermans}(2006)}]{TimmermansPRL2006}
\bibinfo{author}{\bibfnamefont{F.~M.} \bibnamefont{Cucchietti}}
  \bibnamefont{and}
  \bibinfo{author}{\bibfnamefont{E.}~\bibnamefont{Timmermans}},
  \bibinfo{journal}{Phys. Rev. Lett.} \textbf{\bibinfo{volume}{96}},
  \bibinfo{pages}{210401} (\bibinfo{year}{2006}).

\bibitem[{\citenamefont{Blinova et~al.}(2013)\citenamefont{Blinova, Boshier,
  and Timmermans}}]{TimmermansPRA2013}
\bibinfo{author}{\bibfnamefont{A.~A.} \bibnamefont{Blinova}},
  \bibinfo{author}{\bibfnamefont{M.~G.} \bibnamefont{Boshier}},
  \bibnamefont{and}
  \bibinfo{author}{\bibfnamefont{E.}~\bibnamefont{Timmermans}},
  \bibinfo{journal}{Phys. Rev. A} \textbf{\bibinfo{volume}{88}},
  \bibinfo{pages}{053610} (\bibinfo{year}{2013}).

\bibitem[{\citenamefont{Viverit et~al.}(2000)\citenamefont{Viverit, Pethick,
  and Smith}}]{ViveritPRA2000}
\bibinfo{author}{\bibfnamefont{L.}~\bibnamefont{Viverit}},
  \bibinfo{author}{\bibfnamefont{C.~J.} \bibnamefont{Pethick}},
  \bibnamefont{and} \bibinfo{author}{\bibfnamefont{H.}~\bibnamefont{Smith}},
  \bibinfo{journal}{Phys. Rev. A} \textbf{\bibinfo{volume}{61}},
  \bibinfo{pages}{053605} (\bibinfo{year}{2000}).

\bibitem[{\citenamefont{Bijlsma et~al.}(2000)\citenamefont{Bijlsma, Heringa,
  and Stoof}}]{BijlsmaPRA2000}
\bibinfo{author}{\bibfnamefont{M.~J.} \bibnamefont{Bijlsma}},
  \bibinfo{author}{\bibfnamefont{B.~A.} \bibnamefont{Heringa}},
  \bibnamefont{and} \bibinfo{author}{\bibfnamefont{H.~T.~C.}
  \bibnamefont{Stoof}}, \bibinfo{journal}{Phys. Rev. A}
  \textbf{\bibinfo{volume}{61}}, \bibinfo{pages}{053601}
  (\bibinfo{year}{2000}).

\bibitem[{\citenamefont{Santamore and Timmermans}(2011)}]{TimmermansJNP2011}
\bibinfo{author}{\bibfnamefont{D.~H.} \bibnamefont{Santamore}}
  \bibnamefont{and}
  \bibinfo{author}{\bibfnamefont{E.}~\bibnamefont{Timmermans}},
  \bibinfo{journal}{New J. Phys.} \textbf{\bibinfo{volume}{13}}
  (\bibinfo{year}{2011}).

\bibitem[{\citenamefont{Casteels et~al.}(2013)\citenamefont{Casteels, Tempere,
  and Devreese}}]{DevreesePRA2013}
\bibinfo{author}{\bibfnamefont{W.}~\bibnamefont{Casteels}},
  \bibinfo{author}{\bibfnamefont{J.}~\bibnamefont{Tempere}}, \bibnamefont{and}
  \bibinfo{author}{\bibfnamefont{J.~T.} \bibnamefont{Devreese}},
  \bibinfo{journal}{Phys. Rev. A} \textbf{\bibinfo{volume}{88}},
  \bibinfo{pages}{013613} (\bibinfo{year}{2013}).

\bibitem[{\citenamefont{Balewski et~al.}(2013)\citenamefont{Balewski, Krupp,
  Gaj, Peter, B{\"u}chler, L{\"o}w, Hofferberth, and
  Pfau}}]{JonathanNature2013}
\bibinfo{author}{\bibfnamefont{J.~B.} \bibnamefont{Balewski}},
  \bibinfo{author}{\bibfnamefont{A.~T.} \bibnamefont{Krupp}},
  \bibinfo{author}{\bibfnamefont{A.}~\bibnamefont{Gaj}},
  \bibinfo{author}{\bibfnamefont{D.}~\bibnamefont{Peter}},
  \bibinfo{author}{\bibfnamefont{H.~P.} \bibnamefont{B{\"u}chler}},
  \bibinfo{author}{\bibfnamefont{R.}~\bibnamefont{L{\"o}w}},
  \bibinfo{author}{\bibfnamefont{S.}~\bibnamefont{Hofferberth}},
  \bibnamefont{and} \bibinfo{author}{\bibfnamefont{T.}~\bibnamefont{Pfau}},
  \bibinfo{journal}{Nature (London)} \textbf{\bibinfo{volume}{502}},
  \bibinfo{pages}{664} (\bibinfo{year}{2013}).

\bibitem[{\citenamefont{Bardeen et~al.}(1957)\citenamefont{Bardeen, Cooper, and
  Schrieffer}}]{BCStheory}
\bibinfo{author}{\bibfnamefont{J.}~\bibnamefont{Bardeen}},
  \bibinfo{author}{\bibfnamefont{L.~N.} \bibnamefont{Cooper}},
  \bibnamefont{and} \bibinfo{author}{\bibfnamefont{J.~R.}
  \bibnamefont{Schrieffer}}, \bibinfo{journal}{Phys. Rev.}
  \textbf{\bibinfo{volume}{108}}, \bibinfo{pages}{1175} (\bibinfo{year}{1957}).

\bibitem[{\citenamefont{Fermi}(1934)}]{Fermi1934}
\bibinfo{author}{\bibfnamefont{E.}~\bibnamefont{Fermi}}, \bibinfo{journal}{Il
  Nuovo Cimento} \textbf{\bibinfo{volume}{11}}, \bibinfo{pages}{157}
  (\bibinfo{year}{1934}), ISSN \bibinfo{issn}{1827-6121}.

\bibitem[{not({\natexlab{a}})}]{note-p-wave}
\bibinfo{note}{$V_p ({\bf x},{\bf r} ) = (6\pi \hbar^2/m_e) A_p^3(k) \delta
  ^{(3)} \left( {\bf x} - {\bf r} \right) \overleftarrow{\nabla} \cdot
  \overrightarrow{\nabla}$ for $p$-wave, with $k^3 A_p^3(k) = - \tan \delta_p
  (k)$.}

\bibitem[{\citenamefont{Greene et~al.}(2000)\citenamefont{Greene, Dickinson,
  and Sadeghpour}}]{GreenePRL2000}
\bibinfo{author}{\bibfnamefont{C.~H.} \bibnamefont{Greene}},
  \bibinfo{author}{\bibfnamefont{A.~S.} \bibnamefont{Dickinson}},
  \bibnamefont{and} \bibinfo{author}{\bibfnamefont{H.~R.}
  \bibnamefont{Sadeghpour}}, \bibinfo{journal}{Phys. Rev. Lett.}
  \textbf{\bibinfo{volume}{85}}, \bibinfo{pages}{2458} (\bibinfo{year}{2000}).

\bibitem[{\citenamefont{Bendkowsky et~al.}(2009)\citenamefont{Bendkowsky,
  Butscher, Nipper, Shaffer, L{\"o}w, and Pfau}}]{BendkowskyNature2009}
\bibinfo{author}{\bibfnamefont{V.}~\bibnamefont{Bendkowsky}},
  \bibinfo{author}{\bibfnamefont{B.}~\bibnamefont{Butscher}},
  \bibinfo{author}{\bibfnamefont{J.}~\bibnamefont{Nipper}},
  \bibinfo{author}{\bibfnamefont{J.~P.} \bibnamefont{Shaffer}},
  \bibinfo{author}{\bibfnamefont{R.}~\bibnamefont{L{\"o}w}}, \bibnamefont{and}
  \bibinfo{author}{\bibfnamefont{T.}~\bibnamefont{Pfau}},
  \bibinfo{journal}{Nature (London)} \textbf{\bibinfo{volume}{458}},
  \bibinfo{pages}{1005} (\bibinfo{year}{2009}).

\bibitem[{\citenamefont{Li et~al.}(2011)\citenamefont{Li, Pohl, Rost,
  Rittenhouse, Sadeghpour, Nipper, Butscher, Balewski, Bendkowsky, L{\"o}w
  et~al.}}]{LiWscience2011}
\bibinfo{author}{\bibfnamefont{W.}~\bibnamefont{Li}},
  \bibinfo{author}{\bibfnamefont{T.}~\bibnamefont{Pohl}},
  \bibinfo{author}{\bibfnamefont{J.~M.} \bibnamefont{Rost}},
  \bibinfo{author}{\bibfnamefont{S.~T.} \bibnamefont{Rittenhouse}},
  \bibinfo{author}{\bibfnamefont{H.~R.} \bibnamefont{Sadeghpour}},
  \bibinfo{author}{\bibfnamefont{J.}~\bibnamefont{Nipper}},
  \bibinfo{author}{\bibfnamefont{B.}~\bibnamefont{Butscher}},
  \bibinfo{author}{\bibfnamefont{J.~B.} \bibnamefont{Balewski}},
  \bibinfo{author}{\bibfnamefont{V.}~\bibnamefont{Bendkowsky}},
  \bibinfo{author}{\bibfnamefont{R.}~\bibnamefont{L{\"o}w}},
  \bibnamefont{et~al.}, \bibinfo{journal}{Science}
  \textbf{\bibinfo{volume}{334}}, \bibinfo{pages}{1110} (\bibinfo{year}{2011}).

\bibitem[{\citenamefont{Bellos et~al.}(2013)\citenamefont{Bellos, Carollo,
  Banerjee, Eyler, Gould, and Stwalley}}]{BellosPRL2013}
\bibinfo{author}{\bibfnamefont{M.~A.} \bibnamefont{Bellos}},
  \bibinfo{author}{\bibfnamefont{R.}~\bibnamefont{Carollo}},
  \bibinfo{author}{\bibfnamefont{J.}~\bibnamefont{Banerjee}},
  \bibinfo{author}{\bibfnamefont{E.~E.} \bibnamefont{Eyler}},
  \bibinfo{author}{\bibfnamefont{P.~L.} \bibnamefont{Gould}}, \bibnamefont{and}
  \bibinfo{author}{\bibfnamefont{W.~C.} \bibnamefont{Stwalley}},
  \bibinfo{journal}{Phys. Rev. Lett.} \textbf{\bibinfo{volume}{111}},
  \bibinfo{pages}{053001} (\bibinfo{year}{2013}).

\bibitem[{\citenamefont{Krupp et~al.}(2014)\citenamefont{Krupp, Gaj, Balewski,
  Ilzh{\"o}fer, Hofferberth, L{\"o}w, Pfau, Kurz, and
  Schmelcher}}]{Andersonarxiv2014}
\bibinfo{author}{\bibfnamefont{A.}~\bibnamefont{Krupp}},
  \bibinfo{author}{\bibfnamefont{A.}~\bibnamefont{Gaj}},
  \bibinfo{author}{\bibfnamefont{J.}~\bibnamefont{Balewski}},
  \bibinfo{author}{\bibfnamefont{P.}~\bibnamefont{Ilzh{\"o}fer}},
  \bibinfo{author}{\bibfnamefont{S.}~\bibnamefont{Hofferberth}},
  \bibinfo{author}{\bibfnamefont{R.}~\bibnamefont{L{\"o}w}},
  \bibinfo{author}{\bibfnamefont{T.}~\bibnamefont{Pfau}},
  \bibinfo{author}{\bibfnamefont{M.}~\bibnamefont{Kurz}}, \bibnamefont{and}
  \bibinfo{author}{\bibfnamefont{P.}~\bibnamefont{Schmelcher}},
  \bibinfo{journal}{arXiv: 1401.2477v1}  (\bibinfo{year}{2014}).

\bibitem[{\citenamefont{Anderson et~al.}(2014)\citenamefont{Anderson, Miller,
  and Raithel}}]{Krupparxiv2014}
\bibinfo{author}{\bibfnamefont{D.~A.} \bibnamefont{Anderson}},
  \bibinfo{author}{\bibfnamefont{S.~A.} \bibnamefont{Miller}},
  \bibnamefont{and} \bibinfo{author}{\bibfnamefont{G.}~\bibnamefont{Raithel}},
  \bibinfo{journal}{arXiv: 1401.2477v1}  (\bibinfo{year}{2014}).

\bibitem[{\citenamefont{Hamilton et~al.}(2002)\citenamefont{Hamilton, Greene,
  and Sadeghpour}}]{HamiltonJPB2002}
\bibinfo{author}{\bibfnamefont{E.~L.} \bibnamefont{Hamilton}},
  \bibinfo{author}{\bibfnamefont{C.~H.} \bibnamefont{Greene}},
  \bibnamefont{and} \bibinfo{author}{\bibfnamefont{H.~R.}
  \bibnamefont{Sadeghpour}}, \bibinfo{journal}{J. Phys. B: At. Mol. Opt. Phys.}
  \textbf{\bibinfo{volume}{35}}, \bibinfo{pages}{L199} (\bibinfo{year}{2002}).

\bibitem[{\citenamefont{Mayle et~al.}(2012)\citenamefont{Mayle, Rittenhouse,
  Schmelcher, and Sadeghpour}}]{RittenhousePRA2012}
\bibinfo{author}{\bibfnamefont{M.}~\bibnamefont{Mayle}},
  \bibinfo{author}{\bibfnamefont{S.~T.} \bibnamefont{Rittenhouse}},
  \bibinfo{author}{\bibfnamefont{P.}~\bibnamefont{Schmelcher}},
  \bibnamefont{and} \bibinfo{author}{\bibfnamefont{H.~R.}
  \bibnamefont{Sadeghpour}}, \bibinfo{journal}{Phys. Rev. A}
  \textbf{\bibinfo{volume}{85}}, \bibinfo{pages}{052511}
  (\bibinfo{year}{2012}).

\bibitem[{\citenamefont{Karpiuk et~al.}(2014)\citenamefont{Karpiuk, Brewczyk,
  Rza\.{z}ewski, Balewski, Krupp, Gaj, L\"{o}w, Hofferberth, and
  Pfau}}]{PfauArxiv2014}
\bibinfo{author}{\bibfnamefont{T.}~\bibnamefont{Karpiuk}},
  \bibinfo{author}{\bibfnamefont{M.}~\bibnamefont{Brewczyk}},
  \bibinfo{author}{\bibfnamefont{K.}~\bibnamefont{Rza\.{z}ewski}},
  \bibinfo{author}{\bibfnamefont{J.~B.} \bibnamefont{Balewski}},
  \bibinfo{author}{\bibfnamefont{A.~T.} \bibnamefont{Krupp}},
  \bibinfo{author}{\bibfnamefont{A.}~\bibnamefont{Gaj}},
  \bibinfo{author}{\bibfnamefont{R.}~\bibnamefont{L\"{o}w}},
  \bibinfo{author}{\bibfnamefont{S.}~\bibnamefont{Hofferberth}},
  \bibnamefont{and} \bibinfo{author}{\bibfnamefont{T.}~\bibnamefont{Pfau}},
  \bibinfo{journal}{arXiv:1402.6875}  (\bibinfo{year}{2014}).

\bibitem[{\citenamefont{Boisseau et~al.}(2002)\citenamefont{Boisseau, Simbotin,
  and C\^ot\'e}}]{macrodimer-1}
\bibinfo{author}{\bibfnamefont{C.}~\bibnamefont{Boisseau}},
  \bibinfo{author}{\bibfnamefont{I.}~\bibnamefont{Simbotin}}, \bibnamefont{and}
  \bibinfo{author}{\bibfnamefont{R.}~\bibnamefont{C\^ot\'e}},
  \bibinfo{journal}{Phys. Rev. Lett.} \textbf{\bibinfo{volume}{88}},
  \bibinfo{pages}{133004} (\bibinfo{year}{2002}).

\bibitem[{\citenamefont{Lukin et~al.}(2001)\citenamefont{Lukin, Fleischhauer,
  Cote, Duan, Jaksch, Cirac, and Zoller}}]{LukinPRL2001}
\bibinfo{author}{\bibfnamefont{M.~D.} \bibnamefont{Lukin}},
  \bibinfo{author}{\bibfnamefont{M.}~\bibnamefont{Fleischhauer}},
  \bibinfo{author}{\bibfnamefont{R.}~\bibnamefont{Cote}},
  \bibinfo{author}{\bibfnamefont{L.~M.} \bibnamefont{Duan}},
  \bibinfo{author}{\bibfnamefont{D.}~\bibnamefont{Jaksch}},
  \bibinfo{author}{\bibfnamefont{J.~I.} \bibnamefont{Cirac}}, \bibnamefont{and}
  \bibinfo{author}{\bibfnamefont{P.}~\bibnamefont{Zoller}},
  \bibinfo{journal}{Phys. Rev. Lett.} \textbf{\bibinfo{volume}{87}},
  \bibinfo{pages}{037901} (\bibinfo{year}{2001}).

\bibitem[{\citenamefont{Tong et~al.}(2004)\citenamefont{Tong, Farooqi,
  Stanojevic, Krishnan, Zhang, C\^ot\'e, Eyler, and Gould}}]{TongPRL2004}
\bibinfo{author}{\bibfnamefont{D.}~\bibnamefont{Tong}},
  \bibinfo{author}{\bibfnamefont{S.~M.} \bibnamefont{Farooqi}},
  \bibinfo{author}{\bibfnamefont{J.}~\bibnamefont{Stanojevic}},
  \bibinfo{author}{\bibfnamefont{S.}~\bibnamefont{Krishnan}},
  \bibinfo{author}{\bibfnamefont{Y.~P.} \bibnamefont{Zhang}},
  \bibinfo{author}{\bibfnamefont{R.}~\bibnamefont{C\^ot\'e}},
  \bibinfo{author}{\bibfnamefont{E.~E.} \bibnamefont{Eyler}}, \bibnamefont{and}
  \bibinfo{author}{\bibfnamefont{P.~L.} \bibnamefont{Gould}},
  \bibinfo{journal}{Phys. Rev. Lett.} \textbf{\bibinfo{volume}{93}},
  \bibinfo{pages}{063001} (\bibinfo{year}{2004}).

\bibitem[{\citenamefont{Singer et~al.}(2004)\citenamefont{Singer, Reetz-Lamour,
  Amthor, Marcassa, and Weidem\"uller}}]{SingerPRL2004}
\bibinfo{author}{\bibfnamefont{K.}~\bibnamefont{Singer}},
  \bibinfo{author}{\bibfnamefont{M.}~\bibnamefont{Reetz-Lamour}},
  \bibinfo{author}{\bibfnamefont{T.}~\bibnamefont{Amthor}},
  \bibinfo{author}{\bibfnamefont{L.~G.} \bibnamefont{Marcassa}},
  \bibnamefont{and}
  \bibinfo{author}{\bibfnamefont{M.}~\bibnamefont{Weidem\"uller}},
  \bibinfo{journal}{Phys. Rev. Lett.} \textbf{\bibinfo{volume}{93}},
  \bibinfo{pages}{163001} (\bibinfo{year}{2004}).

\bibitem[{\citenamefont{Vogt et~al.}(2006)\citenamefont{Vogt, Viteau, Zhao,
  Chotia, Comparat, and Pillet}}]{VogtPRL2006}
\bibinfo{author}{\bibfnamefont{T.}~\bibnamefont{Vogt}},
  \bibinfo{author}{\bibfnamefont{M.}~\bibnamefont{Viteau}},
  \bibinfo{author}{\bibfnamefont{J.}~\bibnamefont{Zhao}},
  \bibinfo{author}{\bibfnamefont{A.}~\bibnamefont{Chotia}},
  \bibinfo{author}{\bibfnamefont{D.}~\bibnamefont{Comparat}}, \bibnamefont{and}
  \bibinfo{author}{\bibfnamefont{P.}~\bibnamefont{Pillet}},
  \bibinfo{journal}{Phys. Rev. Lett.} \textbf{\bibinfo{volume}{97}},
  \bibinfo{pages}{083003} (\bibinfo{year}{2006}).

\bibitem[{\citenamefont{Liebisch et~al.}(2005)\citenamefont{Liebisch, Reinhard,
  Berman, and Raithel}}]{LiebischPRL2005}
\bibinfo{author}{\bibfnamefont{T.~C.} \bibnamefont{Liebisch}},
  \bibinfo{author}{\bibfnamefont{A.}~\bibnamefont{Reinhard}},
  \bibinfo{author}{\bibfnamefont{P.~R.} \bibnamefont{Berman}},
  \bibnamefont{and} \bibinfo{author}{\bibfnamefont{G.}~\bibnamefont{Raithel}},
  \bibinfo{journal}{Phys. Rev. Lett.} \textbf{\bibinfo{volume}{95}},
  \bibinfo{pages}{253002} (\bibinfo{year}{2005}).

\bibitem[{\citenamefont{Overstreet et~al.}(2009)\citenamefont{Overstreet,
  Schwettmann, Tallant, Booth, and Shaffer}}]{macrodimer-2}
\bibinfo{author}{\bibfnamefont{K.~R.} \bibnamefont{Overstreet}},
  \bibinfo{author}{\bibfnamefont{A.}~\bibnamefont{Schwettmann}},
  \bibinfo{author}{\bibfnamefont{J.}~\bibnamefont{Tallant}},
  \bibinfo{author}{\bibfnamefont{D.}~\bibnamefont{Booth}}, \bibnamefont{and}
  \bibinfo{author}{\bibfnamefont{J.~P.} \bibnamefont{Shaffer}},
  \bibinfo{journal}{Nature Phys.} \textbf{\bibinfo{volume}{5}},
  \bibinfo{pages}{581 } (\bibinfo{year}{2009}).

\bibitem[{\citenamefont{Omont}(1977)}]{Omont1977}
\bibinfo{author}{\bibfnamefont{A.}~\bibnamefont{Omont}}, \bibinfo{journal}{J.
  Phys. (Paris)} \textbf{\bibinfo{volume}{38}}, \bibinfo{pages}{1343}
  (\bibinfo{year}{1977}).

\bibitem[{\citenamefont{Bendkowsky et~al.}(2010)\citenamefont{Bendkowsky,
  Butscher, Nipper, Balewski, Shaffer, L\"ow, Pfau, Li, Stanojevic, Pohl
  et~al.}}]{BendkowskyPRL2010}
\bibinfo{author}{\bibfnamefont{V.}~\bibnamefont{Bendkowsky}},
  \bibinfo{author}{\bibfnamefont{B.}~\bibnamefont{Butscher}},
  \bibinfo{author}{\bibfnamefont{J.}~\bibnamefont{Nipper}},
  \bibinfo{author}{\bibfnamefont{J.~B.} \bibnamefont{Balewski}},
  \bibinfo{author}{\bibfnamefont{J.~P.} \bibnamefont{Shaffer}},
  \bibinfo{author}{\bibfnamefont{R.}~\bibnamefont{L\"ow}},
  \bibinfo{author}{\bibfnamefont{T.}~\bibnamefont{Pfau}},
  \bibinfo{author}{\bibfnamefont{W.}~\bibnamefont{Li}},
  \bibinfo{author}{\bibfnamefont{J.}~\bibnamefont{Stanojevic}},
  \bibinfo{author}{\bibfnamefont{T.}~\bibnamefont{Pohl}}, \bibnamefont{et~al.},
  \bibinfo{journal}{Phys. Rev. Lett.} \textbf{\bibinfo{volume}{105}},
  \bibinfo{pages}{163201} (\bibinfo{year}{2010}).

\bibitem[{not({\natexlab{b}})}]{note-ns}
\bibinfo{note}{This approach is expected to be valid for isolated $n s$ Rydberg
  states considered here \cite{HamiltonJPB2002}.}

\bibitem[{\citenamefont{Singer et~al.}(2005)\citenamefont{Singer, Stanojevic,
  Weidem{\"u}ller, and C{\^o}t{\'e}}}]{SingerJPB2005}
\bibinfo{author}{\bibfnamefont{K.}~\bibnamefont{Singer}},
  \bibinfo{author}{\bibfnamefont{J.}~\bibnamefont{Stanojevic}},
  \bibinfo{author}{\bibfnamefont{M.}~\bibnamefont{Weidem{\"u}ller}},
  \bibnamefont{and}
  \bibinfo{author}{\bibfnamefont{R.}~\bibnamefont{C{\^o}t{\'e}}},
  \bibinfo{journal}{J. Phys. B: At. Mol. Opt. Phys.}
  \textbf{\bibinfo{volume}{38}}, \bibinfo{pages}{S295} (\bibinfo{year}{2005}).

\bibitem[{\citenamefont{Roberts and Rica}(2009)}]{RobertsPRL2009}
\bibinfo{author}{\bibfnamefont{D.~C.} \bibnamefont{Roberts}} \bibnamefont{and}
  \bibinfo{author}{\bibfnamefont{S.}~\bibnamefont{Rica}},
  \bibinfo{journal}{Phys. Rev. Lett.} \textbf{\bibinfo{volume}{102}},
  \bibinfo{pages}{025301} (\bibinfo{year}{2009}).

\bibitem[{\citenamefont{Osychenko et~al.}(2012)\citenamefont{Osychenko,
  Astrakharchik, Mazzanti, and Boronat}}]{OsychenkoPRA2012}
\bibinfo{author}{\bibfnamefont{O.~N.} \bibnamefont{Osychenko}},
  \bibinfo{author}{\bibfnamefont{G.~E.} \bibnamefont{Astrakharchik}},
  \bibinfo{author}{\bibfnamefont{F.}~\bibnamefont{Mazzanti}}, \bibnamefont{and}
  \bibinfo{author}{\bibfnamefont{J.}~\bibnamefont{Boronat}},
  \bibinfo{journal}{Phys. Rev. A} \textbf{\bibinfo{volume}{85}},
  \bibinfo{pages}{063604} (\bibinfo{year}{2012}).

\end{thebibliography}
\end{document}